\documentclass[fleqn,usenatbib]{mnras}

\usepackage{newtxtext,newtxmath}

\usepackage[T1]{fontenc}

\DeclareRobustCommand{\VAN}[3]{#2}
\let\VANthebibliography\thebibliography
\def\thebibliography{\DeclareRobustCommand{\VAN}[3]{##3}\VANthebibliography}

\usepackage{graphicx}	
\usepackage{amsmath}	
\usepackage[normalem]{ulem}
\usepackage{dirtytalk}
\usepackage{hyperref}
\usepackage[dvipsnames]{xcolor}
\usepackage[nameinlink, capitalise, noabbrev]{cleveref}

\title[Radio AGN fraction around clusters]{From outskirts to core: the suppression and activation of radio AGN around galaxy clusters}

\author[K. de Vos et al.]{
K. de Vos,$^{1}$\thanks{E-mail: ppykd1@nottingham.ac.uk (KdV)}
N. A. Hatch,$^{1}$
M. R. Merrifield,$^{1}$
\\
$^{1}$School of Physics and Astronomy, University Park, University of Nottingham, Nottingham, NG7 2RD, UK}

\date{Accepted XXX. Received YYY; in original form ZZZ}

\pubyear{2024}

\begin{document}
\label{firstpage}
\pagerange{\pageref{firstpage}--\pageref{lastpage}}
\maketitle

\begin{abstract}
To investigate how the radio-identified active galactic nuclei (AGN) fraction varies with cluster-centric radius, we present the projected and de-projected distributions of a large sample of LOFAR-identified radio AGN out to $30R_{500}$ around galaxy clusters. The AGN fraction experiences a $\sim 25\%$ increase above the field fraction in the cluster outskirts at around $10R_{500}$, a $\sim 20\%$ decrease around $\sim 0.5R_{500}$, and an increase of over three times the field fraction value in the very cluster core. We label these three radial windows the \textit{outer, intermediate} and \textit{inner} regions respectively, and investigate how these radial trends might arise due to intrinsic properties of the AGN population. The only difference seen in host galaxy stellar mass is in the \textit{inner} region, where there is a much higher fraction of massive host galaxies. Analysing AGN radio luminosity, regions with a higher AGN fraction tend to have more radio luminous AGN, and vice versa. We discuss the physical mechanisms that might be responsible for these results with reference to the literature.

\end{abstract}




\section{Introduction}
\label{sec:Introduction}

It is well established that both internal feedback from active galactic nuclei \citep[AGN,][]{Kauffmann2003, Best2005a, Schawinski2007, Hickox2009, Best2012} and external environmental influences \citep{Kauffmann2004, Park2007, Silverman2008, Ellison2011, Sabater2013} play  key roles in shaping the properties of galaxies.  However, if AGN activity is, in turn, affected by the wider environment, then these two factors are not entirely independent, and study of the coupling between them is required to fully understand the drivers of galaxy evolution.

Since nuclear activity depends on the abundance of available gas to feed the AGN, which can be affected by wider environmental factors \citep{Gunn1972, Larson1980, Farouki1981, Moore1996, Makino1997, Abadi1999}, and the triggering of the AGN also depends on external phenomena that might disturb gas orbits to funnel material to the galactic centre \citep{Best2007, Poggianti2017, Marshall2018, Ricarte2020, Peluso2022}, such environmental influences on AGN activity might be expected.  Indeed, a recent large-scale simulation incorporating these physical processes \citep{Rihtarsic2023} confirmed that the prevalence of AGN does depend significantly on distance from the nearest cluster of galaxies, with an enhancement in activity above the field level out to surprisingly large radii, and a suppression of activity within the cluster itself, apart from a spike in AGN activity at the very centre of the cluster.

However, observational support for this expectation is rather mixed. Many studies have found little if any variation with distance from a cluster apart from an increase in its very central region \citep{Ruderman2005, Gilmour2009, Manzer2014, Mo2018, Hashiguchi2023}. This central excess seems to be attributable to brightest cluster galaxies (BCGs) hosting a higher fraction of AGN \citep{Best2007}.  Other works have found that, far from increasing, the prevalence of AGN activity seems to be somewhat suppressed in the regions around clusters \citep{Ehlert2013, Ehlert2014, Gordon2018, Koulouridis2024}.

The reason for the lack of conclusive observations seems to be that the effects are quite modest in amplitude, and the large homogeneous data set required to measure them unequivocally has not thus-far been available.  Fortunately, the combination of the Low-Frequency Array (LOFAR) Two-Metre Sky Survey (LoTSS) second data release \citep[DR2,][]{Shimwell2022} with the Sloan Digital Sky Survey (SDSS) sixteenth data release \citep[DR16,][]{Ahumada2020} provides just such a data set: its area allows the distribution of a very large number of galaxies around clusters to be studied in detail, and a well-defined sample of AGN can reliably be identified from their radio emission.  In this paper, we use these data to determine how the fraction of AGN varies with position out to large radii around clusters.

\section{Data and Method}
\label{sec:Data_Method}

\subsection{Data and sample selection}
\label{subsec:data_sample_selec}

We primarily follow the data and methods set out in \citet{devos2024}, which is summarised below. In this work we reference $R_{500}$, which is defined as the cluster radius within which the average density is five hundred times the critical density of the universe, $\rho_c$, where $R_{500}\sim 0.7R_{200}$ \citep{Navarro1996}. $\sigma_{500}$ is the characteristic velocity of the cluster within this radius.

We identify a large sample of 68\,168 radio galaxies (hereafter referred to as \textit{radio galaxies}) from the Low-Frequency Array (LOFAR) Two-metre Sky Survey (LoTSS) second data release \citep[DR2,][]{Shimwell2022}, which are matched to galaxies from the Sloan Digital Sky Survey (SDSS) sixteenth data release \citep[DR16,][]{Ahumada2020} within 2 arcsecs on the sky, and complemented with further data from the MPA-JHU value-added catalogue \citep{Brinchmann2004} and the Wide-field Infrared Survey Explorer (WISE) allWISE source catalogue \citep{Wright2010}.

We also create a catalogue of over 3 million optically-identified galaxies with spectroscopic redshifts from SDSS DR16 (hereafter referred to as \textit{SDSS galaxies}, complemented by both MPA-JHU and WISE data, in order to have a total galaxy population sample with which we can compare the radio galaxy sample. Furthermore, we utilise the cluster catalogues produced by \citet{Wen2012} and \citet{Wen2015} in order to determine the environments of the samples (see \cref{subsec:cluster_matching}), keeping only the sources with spectroscopic redshifts to reduce line-of-sight contamination when matching galaxies and clusters. Further details on the specifics of these surveys and how they are matched and joined in order to create the samples used in this work are given in \citet{devos2024}.

We calculate a stellar mass proxy, $M_{i}$, for the galaxies in these samples, as only 23\% of the sources in \textit{SDSS galaxies} have good stellar mass values from MPA-JHU. We define this mass proxy using the i-band fluxes, $i$, of the galaxies, using the following best fit formula:

\begin{equation}
    \log_{10}(M_{i}) = -0.42i + 1.93.
\end{equation}

Radio luminosities, $L_{150}$, are calculated using the formula
\begin{equation}
    L_{150} = \frac{S_{\textrm{obs}}4 \pi {D_L}^2}{(1+z)^{1+\alpha}},
\end{equation}
where $S_{\rm{obs}}$ is the observed flux density in W\,m$^{-2}$\,Hz$^{-1}$, $D_L$ is the luminosity distance in metres, $z$ is the redshift, and we take the spectral index to be $\alpha=-0.7$ \citep{Shimwell2017}.

We limit \textit{SDSS galaxies}, \textit{radio galaxies} and the cluster catalogues to the same region of sky before matching, and restrict them to the redshift range $0.05 < z < 0.2$. The lower limit is implemented to ensure that single galaxies are not resolved into multiple sources at low redshifts, and the upper limit is chosen in order to reduce any evolutionary effects due to redshift. To test and confirm that there is no redshift dependency, we bin our sample into smaller redshift ranges and find no significant differences between their results and trends.

\subsection{Classification of AGN}
\label{subsec:AGN_classification}

Radio luminosity may be present due to star-formation rather than from radio-loud AGN. In order to separate these star-forming galaxies (SFGs) from the AGN in \textit{radio galaxies}, we use the Wide-field Infrared Survey Explorer (WISE) \citet{Wright2010} colour diagnostic cut: $W2 - W3 = 0.8$, presented in \citet{Sabater2019}, which results in an AGN sample of 13\,053 sources, and is hereafter referred to as \textit{radio AGN}.

We choose this diagnostic for several reasons. Firstly, \citet{Sabater2019} show this classification method to be surprisingly robust when overlaying the final combined classification of each source onto each diagnostic diagram. Secondly, emission line data is not available from MPA-JHU for every radio source in our sample, whereas WISE data is, and so we opt to primarily use the WISE colour diagnostic in order to maintain as large a sample as possible, whilst still retaining a good level of purity.

However, in order to check that the results found in this paper are not due to this choice of AGN classification, we take a smaller test sample that satisfies not just the WISE colour diagnostic, but also the \say{$D_{4000}$ vs. $L_{150MHz}/M_{*}$} method developed by \citet{Best2005b}, and the \say{BPT} diagnostic developed by \citet{Baldwin1981}, both of which are also presented in \citet{Sabater2019}. We find that the results with this smaller, stricter sample are largely consistent with the results seen from the purely WISE colour-cut sample, and thus we can be confident that any trends are not due to the AGN diagnostic method.

After the AGN sample is defined, we determine a 95\% mass threshold cut from \textit{radio AGN} in redshift bins of width 0.01, which we then apply to both \textit{radio AGN} and \textit{SDSS galaxies}, in order to ensure that both samples are directly comparable in their mass limits at all redshifts. This cut corresponds to a lower mass limit range of (0.2--1)$\times 10^{11} M_\odot$ across the redshift range of the sample, resulting in a final sample size of 12\,380 \textit{radio AGN}, and 141\,586 \textit{SDSS galaxies}.

\subsection{Cluster-galaxy association}
\label{subsec:cluster_matching}

In order to begin the process of associating galaxies with clusters, both the \textit{SDSS galaxies} and \textit{radio AGN} samples are matched to a catalogue of 158\,103 clusters from the \citet{Wen2012} and \citet{Wen2015} cluster catalogues after they have undergone the sky and redshift cuts described in \cref{subsec:data_sample_selec}. We do this matching process by following the method described in \citet{devos2024} (adapted from \citet{Garon2019}), which matches galaxies to the cluster with which they are most likely to be associated at redshift $z_{clus}$ via the formula
\begin{equation}
    \frac{|z_{gal}-z_{clus}|}{1+z_{gal}}<0.04,
    \label{eq:z_diff}
\end{equation}
where $z_{gal}$ is the redshift of the galaxy. These galaxies are then restricted on the plane of the sky by only keeping those within the radially projected distance of $<50r_{500}$. This large radius means that some galaxies are matched to multiple clusters, as at this early stage it is difficult to determine with which cluster individual objects are mostly closely associated. However, we attempt to resolve this issue statistically in \cref{subsec:bg_subtraction}.

We have carried out this analysis without removing duplicates, so that a galaxy can be associated with more than one cluster. It is unclear whether this is entirely correct -- if a galaxy is very close to one cluster, it seems unlikely that a more distant neighbour would have any significant impact, but an object roughly mid-way between two clusters could fall under the influence of either or both. To test the significance of this ambiguity, we repeated the analysis with each galaxy only associated with its closest cluster. Fortunately, the number of such duplicates is sufficiently small that the results were unchanged to within their errors by this more stringent selection, so we can conclude that this issue is not a major concern.

\subsection{Background subtraction}
\label{subsec:bg_subtraction}

Due to the relatively large $\Delta z \sim 0.04(1+z)$ used to match galaxies to clusters, it is likely that we include non-associated field galaxies, which we remove statistically using the method described below.

\begin{figure}
    \centering
    \includegraphics[width=\columnwidth]{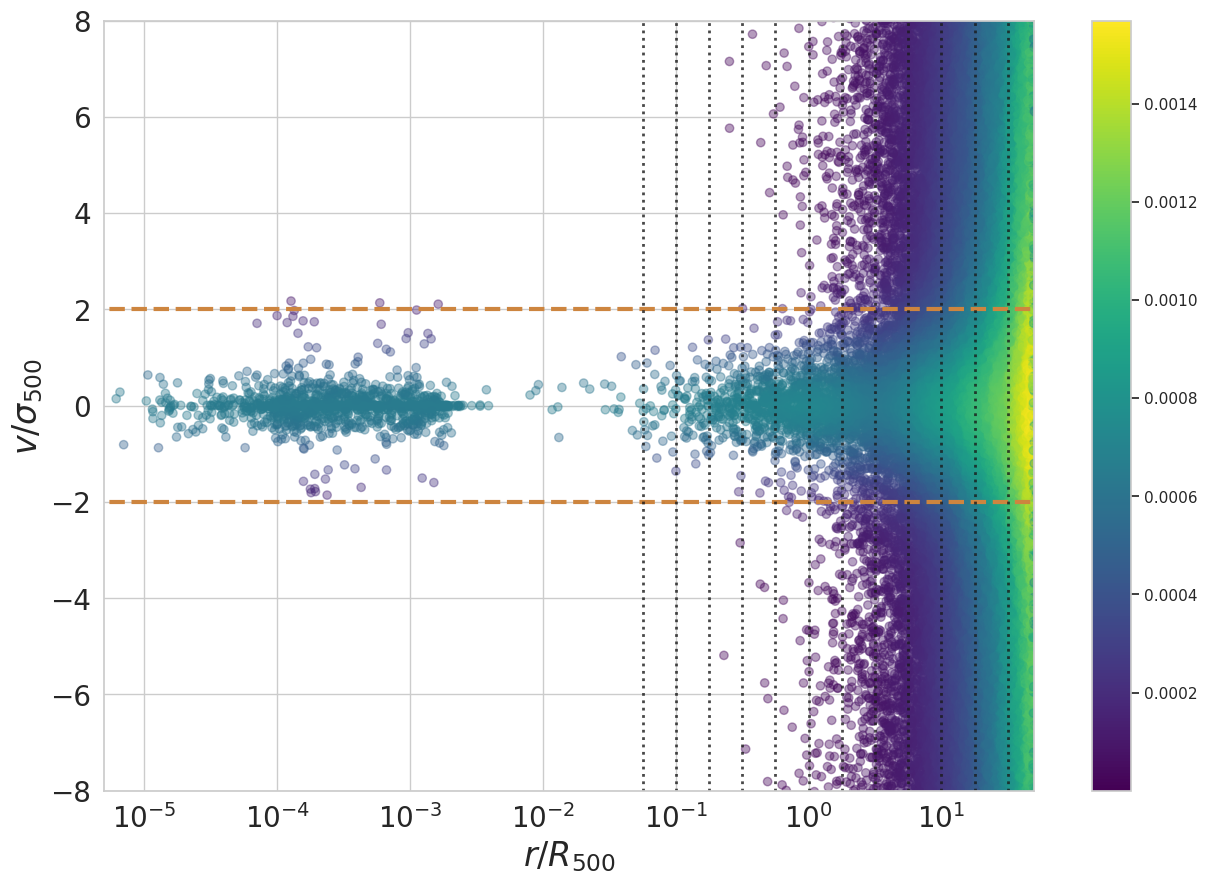}
    \caption{The projected phase-space distribution (showing line-of-sight velocity versus projected separation) for \textit{radio AGN} out to $50r_{500}$. The orange, horizontal, dashed lines at $v = \pm 2\sigma_{500}$ indicate the limits of velocity associated with the cluster, and the black, vertical, dotted lines indicate the cluster radius slicing used to determine the line-of-sight contamination of field galaxies in each annulus bin. The variation in colour amongst the scatter shows the Gaussian kernel density estimation, the values of which are given by the colour bar.}
    \label{fig:phase_space}
\end{figure}

\begin{figure}
    \centering
    \includegraphics[width=\columnwidth]{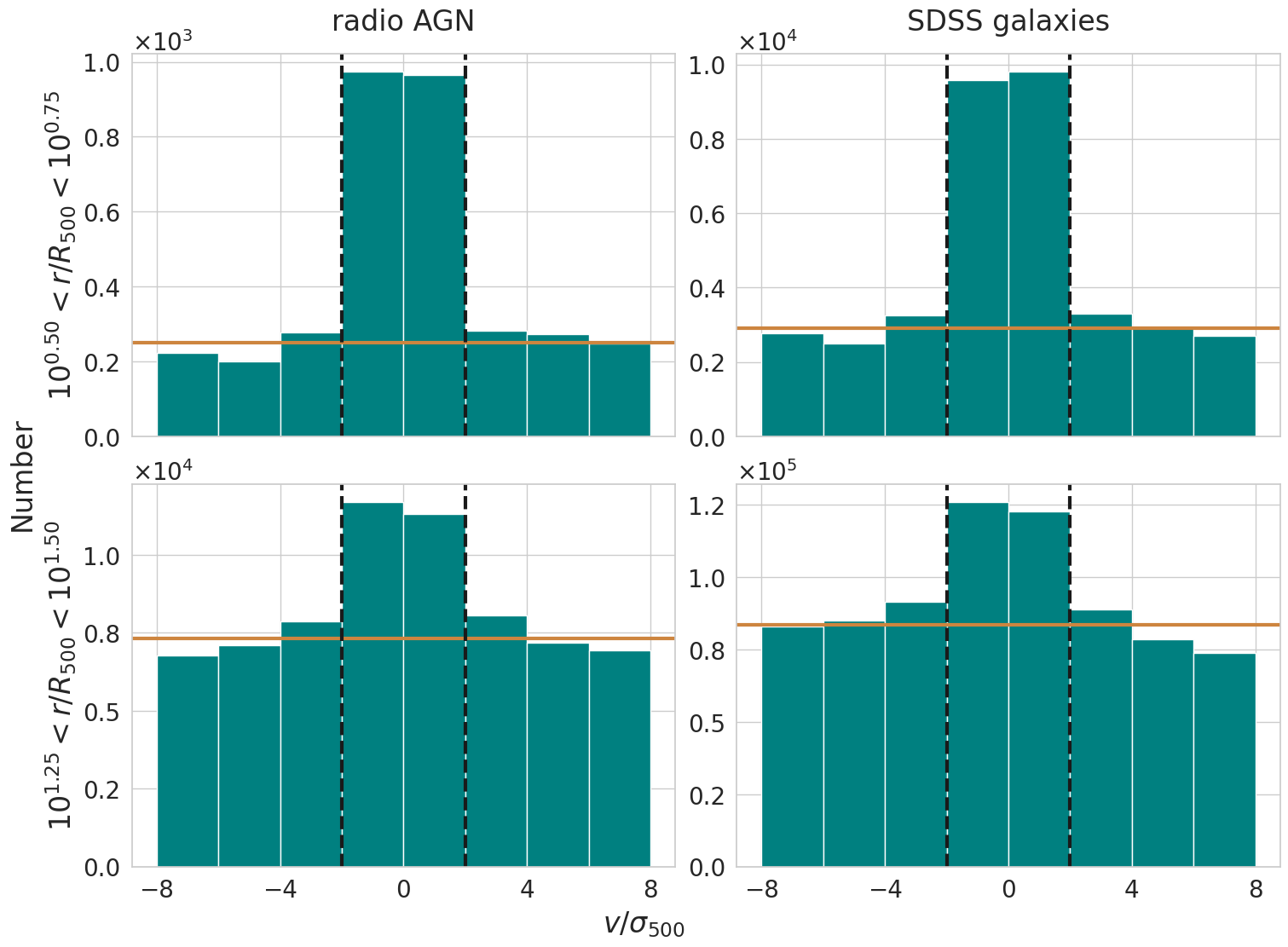}
    \caption{The line-of-sight velocity of galaxies within two of the logarithmically-spaced annuli from the phase spaces of \textit{radio AGN} (left) and \textit{SDSS galaxies} (right). Galaxies within the black, vertical, dashed lines (indicating the characteristic velocity range $v = \pm 2\sigma_{500}$) are deemed to be correctly associated with their host cluster, and those outside are identified as field galaxies. The orange, horizontal, solid line shows the mean of the background contamination in each bin.}
    \label{fig:bg_hists}
\end{figure}

This process is done by first normalising each galaxy's distance from its host cluster, $r$, by that cluster's $R_{500}$ to give $r/R_{500}$, and its associated velocity, $v$ by the cluster's characteristic velocity $\sigma_{500}$ to give $v/\sigma_{500}$, so that data from all clusters in the sample can be stacked. We can then analyse the phase spaces of the two, cluster-matched samples \textit{SDSS galaxies} and \textit{radio AGN} individually, as shown in \cref{fig:phase_space}, in order to statistically determine how many galaxies are associated with clusters at varying annuli. We take logarithmically-spaced annulus bins in $r/R_{500}$ and plot the histograms of $v/\sigma_{500}$ for the galaxies within each bin, in order to determine the signal of cluster-associated galaxies, and contamination of field galaxies, within each annulus, as shown in \cref{fig:bg_hists}.

We take galaxies within a velocity dispersion window of $v = \pm 2\sigma_{500}$ to be associated with their respective clusters, and those outside to be field galaxies which we dub \say{background contamination}, as they are contaminating the sample we consider to be associated with clusters. Thus, in order to retrieve a background-subtracted signal in each annulus, we take the mean of the background contamination and subtract it from the cluster-associated galaxies within $v = \pm 2\sigma_{500}$. Finally, the background-subtracted signal is divided by the annulus area in order to give the number density, $N(R)$, for said annulus, and we derive the error on the number density in each annulus from Poisson statistics. The number of SDSS field galaxies, $N_{SDSS, f}$, and radio field galaxies, $N_{AGN, f}$, are calculated by summing the number of galaxies outside of the $v = \pm 2\sigma_{500}$ in every annulus, for the \textit{SDSS galaxies} and \textit{radio AGN} samples respectively.

\subsection{Abel inversion}
\label{subsec:abel_inversion}

As the number densities calculated in \cref{subsec:bg_subtraction} will only show us a projected distribution with respect to cluster radius, which dilutes the intrinsic cluster properties, we must de-project these values into volume densities in order to get an idea of the true radial galaxy distribution. We can assume spherical symmetry during this process, as the act of stacking multiple clusters together averages out any asymmetries in structure. This assumption allows us to use a simple Abel inversion to convert projected densities into spatial densities, using the formula

\begin{equation}
    \label{eq:Abel_Inv}
    n(r) = -\frac{1}{\pi}\int_{r}^{\infty}\frac{dN}{dR}\frac{1}{(R^2-r^2)^{\frac{1}{2}}}dR,
\end{equation}

where $n(r)$ and $r$ are the de-projected number density and radial distance respectively, and $N(R)$ and $R$ are the projected equivalent variables.

The nature of the numerical implementation of this method means that it is not possible to calculate a spatial density value for the outermost radial bin. Fortunately, we can substitute in the overall field value calculated from \cref{subsec:bg_subtraction} at this radius instead, which acts as a boundary condition and helps suppress the inherent error-amplification seen in such de-projections. Finally, we calculate errors on these volume densities by implementing a Monte Carlo method that resamples the projected data using their Poisson errors. 

We show in \citet{devos2024} that this Abel inversion method agrees with the number density profiles seen in \citet{Beers1986}; we direct the reader towards that publication if they would like a more detailed description of the method described above.

\section{Results}
\label{sec:Results}

\subsection{Total AGN fraction}
\label{subsec:total_AGN_frac}

\begin{figure}
    \centering
    \includegraphics[width=\columnwidth]{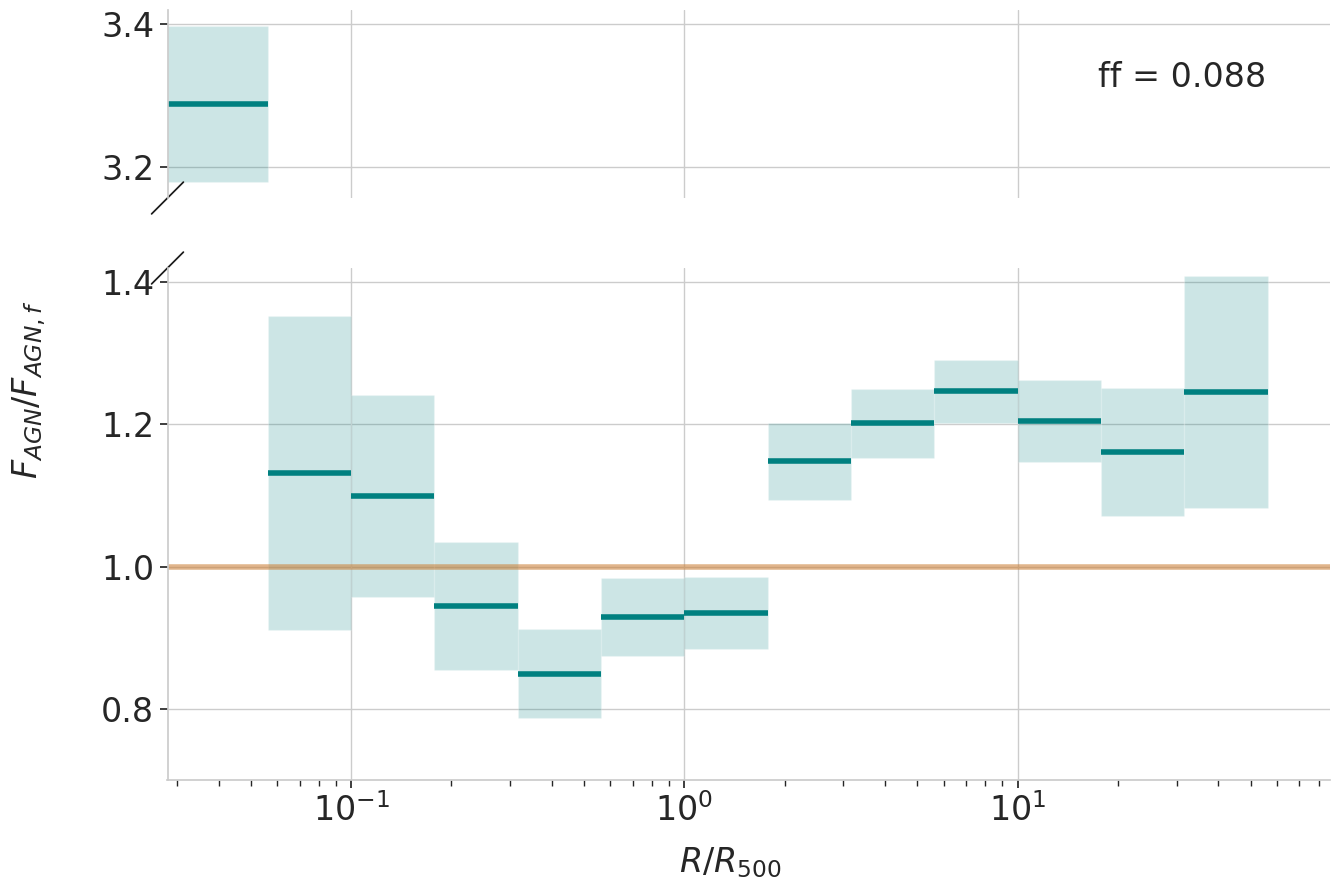}
    \caption{The projected distribution of the relative fraction of galaxies identified as active, with respect to projected cluster radius. The fraction values have been normalised with respect to the AGN field fraction, which is $\sim 0.09$. As such, the orange, horizontal, solid line marks the normalised AGN fraction in the field, which is $1$. The innermost bin includes all galaxies between $0 < R < 10^{-1.25}R_{500}$.}
    \label{fig:AGN_frac}
\end{figure}

With use of the two samples outlined above: \textit{SDSS galaxies} and \textit{radio AGN}, we are able to determine how the AGN fraction varies with respect to projected cluster-centric radius. We calculate this fraction in each annulus using the formula $F_{AGN}=N_{AGN}/N_{SDSS}$, where $N_{AGN}$ and $N_{SDSS}$ are the number densities of \textit{radio AGN} and \textit{SDSS galaxies} respectively. We choose to merge all galaxies into one bin at $R < 10^{-1.25}R_{500}$ due to the increasingly small number of AGN ($<10$ per bin) in the individual logarithmically-spaced annuli within this radius. Finally, in order to more quantitatively define any deviation from the field fraction in the distribution, we normalise $F_{AGN}$ by the field fraction, $F_{AGN,f}$, which is found to be $\sim 0.09$. We calculate the field AGN fraction by taking $F_{AGN,f} = N_{AGN,f}/N_{SDSS,f}$. The resulting cluster-centric distribution can be seen in \cref{fig:AGN_frac}.

\begin{figure}
    \centering
    \includegraphics[width=\columnwidth]{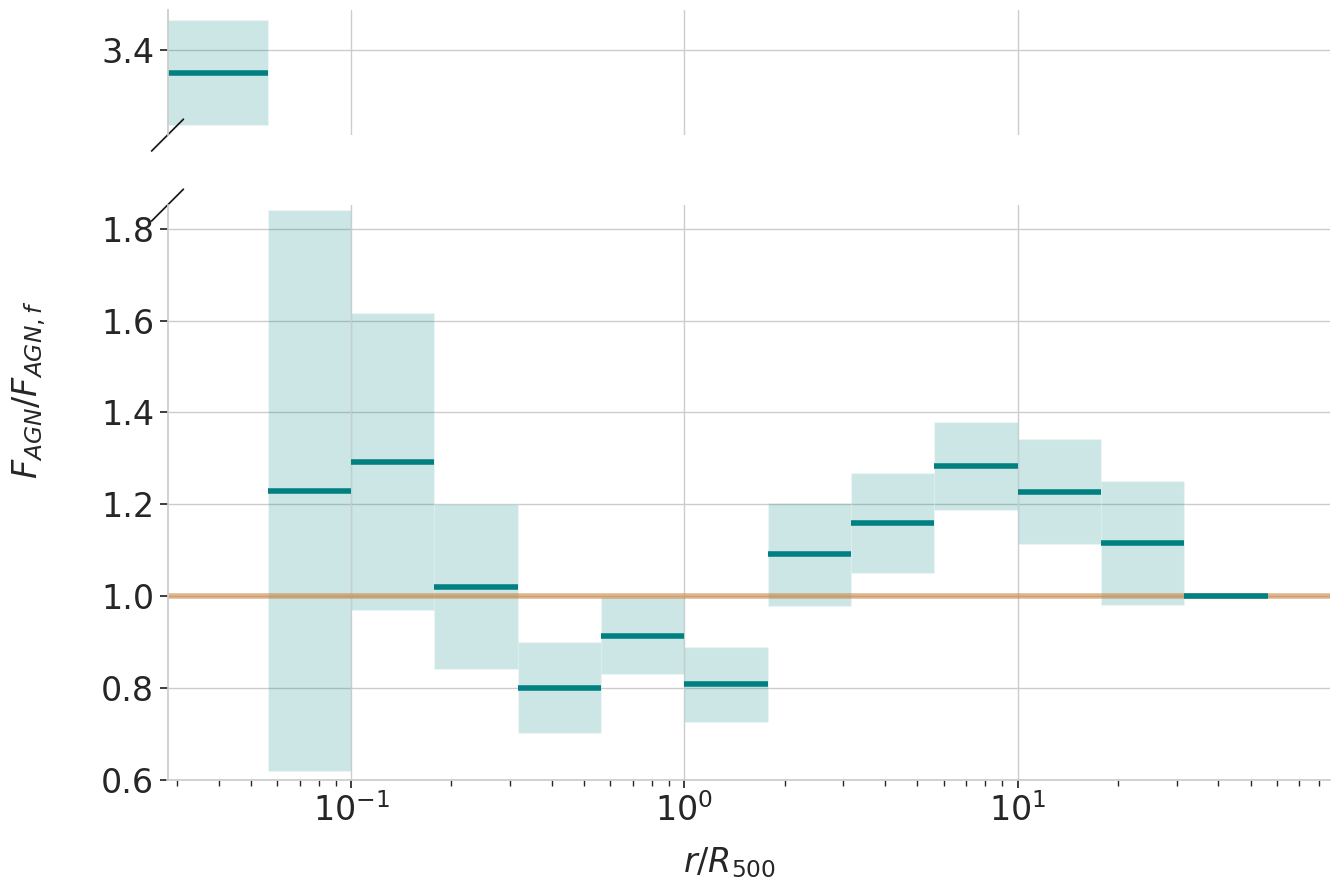}
    \caption{The de-projected distribution of \cref{fig:AGN_frac}, showing the relative fraction of galaxies identified as AGN, with respect to cluster radius.}
    \label{fig:AGN_frac_AI}
\end{figure}

This figure, however, is a projection of the true radial trend, meaning each annulus has the effect of combining data from a range of radii. This plot will therefore tend to average away some of the true variation with radius, and so we de-project the distribution using the Abel inversion method described in \cref{subsec:abel_inversion}. In \cref{fig:AGN_frac_AI} we show the de-projected distribution of the trend seen in \cref{fig:AGN_frac}.

From both of these figures, it is immediately apparent that there is a trend resembling a \say{wiggle} that has not been seen before with observational data. We see that out at around $10R_{500}$, the AGN fraction has risen above the field fraction to peak at a $\sim 25\%$ increase, with a $5.5\sigma$ significance for the projected distribution in \cref{fig:AGN_frac}, and a $3\sigma$ significance for the de-projection in \cref{fig:AGN_frac_AI}. The AGN fraction then decreases with decreasing cluster-centric radius, until it reaches a minimum of $\sim 20\%$ below the field fraction line at around $0.5R_{500}$, with a $2.4\sigma$ significance in \cref{fig:AGN_frac}, and a $2\sigma$ significance in \cref{fig:AGN_frac_AI}. The distribution finally trends back up again in the centre of the cluster, and culminates in a large spike at the very core, with a fraction over three times more than the field fraction. In order to test the significance of the features in this trend against a uniform distribution centred at 1, we perform a $\chi^2$ Test with the projected distribution data, not including the innermost bin, which results in a significance of more than 99.99 per cent for $\nu=10$, where $\nu$ is the number of annulus bins minus the number of model parameters. It is worth noting that the qualitative shape of this variation with radius is very similar to that seen in \citet{Rihtarsic2023}; we discuss the implications of this similarity in \cref{sec:Discussion}.

To explore the possible cause of these features, we split the cluster-centric radial bins into three regions: the \textit{outer} region, defined by a radius of $r > 10^{0.25}R_{500}$ ($\sim 2R_{500}$); the \textit{intermediate} region, given by $10^{-0.75}R_{500} < r < 10^{0.25}R_{500}$; and the \textit{inner} region, for $r < 10^{-0.75}R_{500}$ ($\sim 0.2R_{500}$). These three regions are simply dictated by the radii at which the distribution crosses the field fraction line. This split allows us to investigate if the AGN in these three regions have different intrinsic properties or host galaxies that might be the cause of the fractional variation we are seeing as a function of cluster-centric distance.

\subsection{Mass distribution}
\label{subsec:mass_func}

\begin{figure}
    \centering
    \includegraphics[width=\columnwidth]{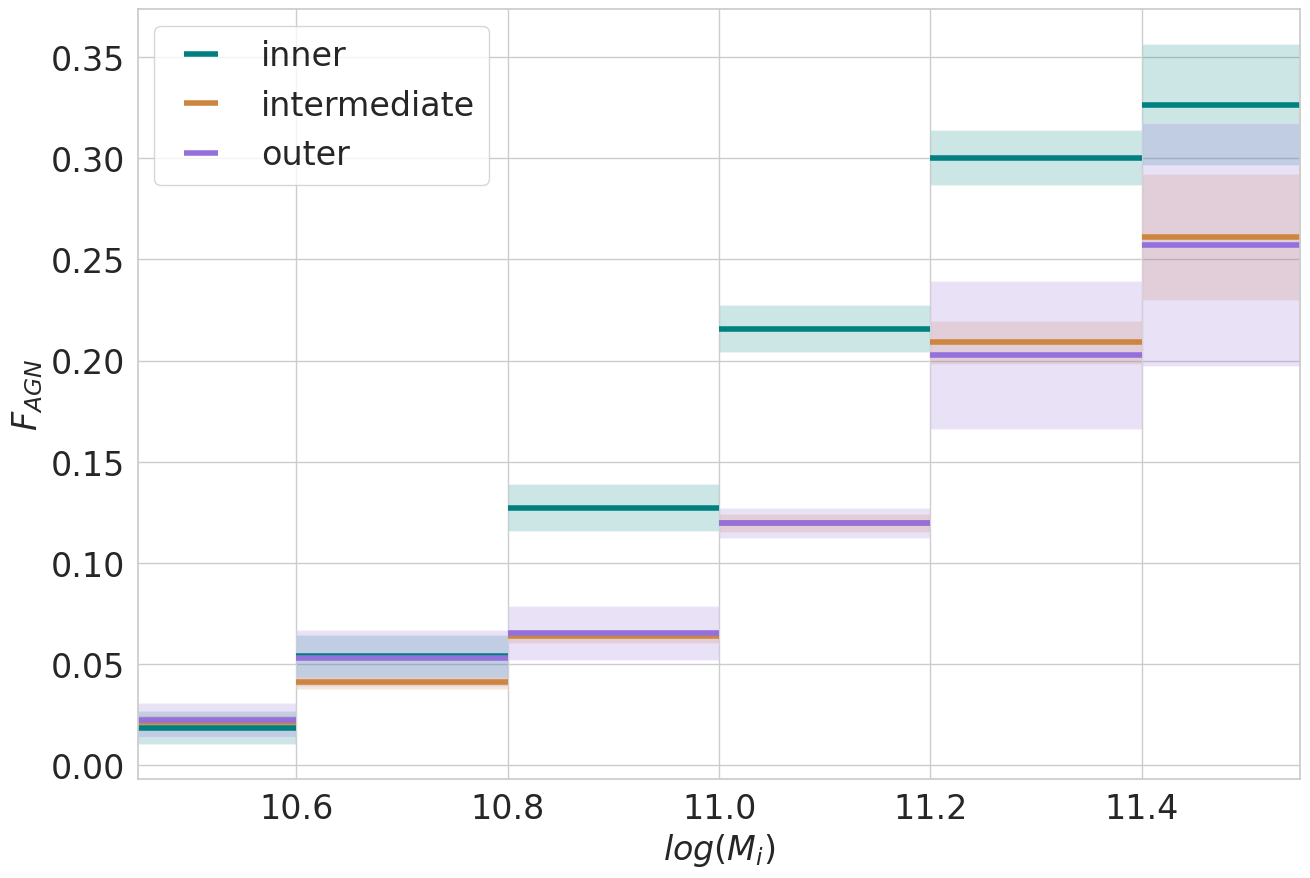}
    \caption{The i-band stellar mass distribution of the host galaxies of $F_{AGN}$ separated into three distinct regions: \textit{inner}, \textit{outer} and \textit{intermediate}. These regions are defined by the radii at which the AGN fraction distributions in \cref{fig:AGN_frac,fig:AGN_frac_AI} cross the field line. }
    \label{fig:Mass_func}
\end{figure}

We start by analysing the stellar mass distribution of the AGN host galaxies for these three regions, as shown in \cref{fig:Mass_func}. Here we can see that, for all three radial ranges, there is an increase in the AGN fraction with mass, as expected \citep{Sabater2019}. However, it is not the absolute values that are informative here, but rather it is the differential between the regions that provides us with information that might suggest that one region has different intrinsic properties to the norm.

For instance, it is immediately apparent that there is a much higher fraction of galaxies in the \textit{inner} region that have high stellar mass, which is to be expected as this region includes massive BCGs that host many AGN. There is no statistically significant difference between the \textit{intermediate} and \textit{outer} region mass distributions, which brings us to the conclusion that the stellar mass of the host galaxy is not the driver behind the fractional AGN differences seen in \cref{fig:AGN_frac,fig:AGN_frac_AI} between these two regions.

\subsection{Radio luminosity}
\label{subsec:radio_lum}

\begin{figure}
    \centering
    \includegraphics[width=\columnwidth]{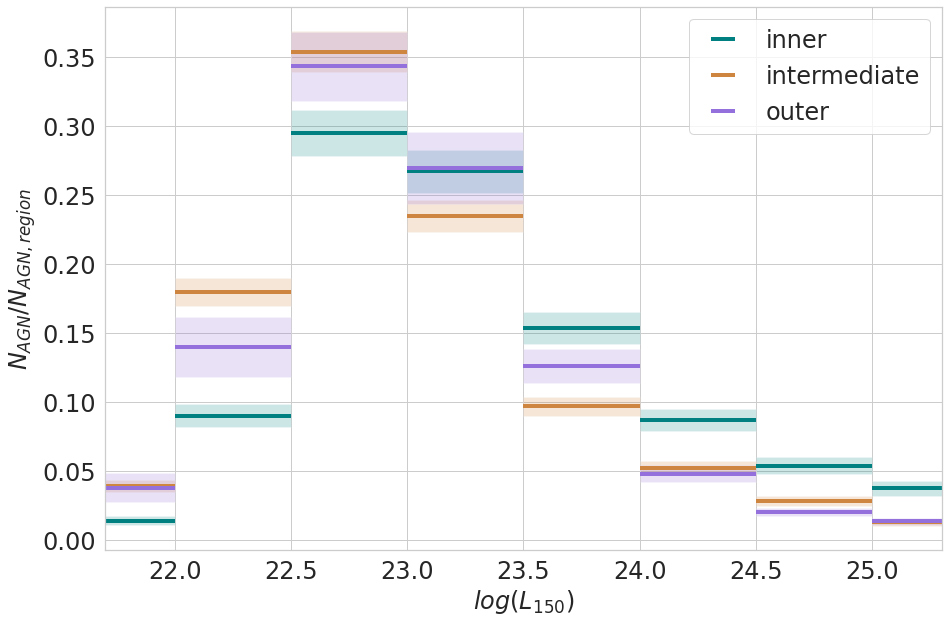}
    \caption{The $L_{150}$ luminosity pdf of the \textit{radio AGN} sample separated into three distinct regions: \textit{inner}, \textit{outer} and \textit{intermediate}. These regions are defined by the radii at which the AGN fraction distributions in \cref{fig:AGN_frac,fig:AGN_frac_AI} cross the field line.}
    \label{fig:Lum_func}
\end{figure}

To assess whether the nature of AGN activity varies with radius, we next consider radio luminosity, $L_{150}$, which is investigated using the radio luminosity probability density function (pdf) in \cref{fig:Lum_func}. Here, as \textit{SDSS galaxies} cannot be binned by $L_{150}$, we present $N_{AGN}/N_{AGN,region}$ on the y-axis, which is the number of AGN in a given region and luminosity bin, normalised by the total number of AGN in that region.

Looking at the differential between the three regions, we find that AGN in the \textit{inner} region have a higher luminosity on average than the other two regions. This result, again, is to be expected due to the BCGs in this sub-sample, which are known for both being the most massive galaxies at a given redshift, and for having higher radio luminosities for a given stellar mass \citep{VonDerLinden2007, Best2007}. We do notice, however, that the \textit{intermediate} region has the highest fraction of low-luminosity AGN, and generally the lowest fraction of high-luminosity AGN, making the \textit{outer} region the middling luminosity sample on average; this finding indicates that the nature of AGN activity does depend on cluster-centric radius.

\section{Summary \& Discussion}
\label{sec:Discussion}

In this paper, we have seen from both \cref{fig:AGN_frac,fig:AGN_frac_AI} that the distribution of the fraction of radio AGN with respect to cluster-centric radius takes the form of a \say{wiggle}. In the \textit{outer} regions, past $\sim 2R_{500}$, we see an increase of up to $\sim 25\%$ above the field fraction, followed by a decrease of $\sim 20\%$ relative to the field in the \textit{intermediate} region, and a huge spike in the very core, \textit{inner} region of the cluster. Although they explored a smaller range of radii, the shape of this distribution is very similar to that seen for massive, quiescent AGN in the recent simulation paper from \citet{Rihtarsic2023}, although we note that the effects were not seen on quite the same scale that we have found observationally. In their paper, they propose that the surge of excess AGN in the outskirts of the cluster is due to the velocity dispersion being lower in this region, thus allowing for a greater frequency of mergers that trigger AGN activity, which would be in agreement with other findings in the literature \citep{Ruderman2005, Fassbender2012, Haines2012, Koulouridis2019}. They also postulate that the reduction seen in the \textit{intermediate} region might be due to the opposite effect, of less mergers taking place due to the higher velocity dispersion in the cluster itself. The spike of radio AGN seen in the cluster core can be explained by the work of \citet{Best2007}, which shows that the accretion of hot gas from a strong cooling flow increases the likelihood that BCGs host radio AGN.

This scenario is consistent with the luminosity variations seen in the three populations observed in \cref{fig:Lum_func}. Given that the fraction of AGN is declining in the \textit{intermediate} region, it would stand to reason that the AGN populations in this area may be starved of fuel due to less frequent mergers, resulting in a lower radio luminosity. Similarly, the potential mergers providing fuel and generating the excess of AGN in the \textit{outer} regions may be responsible for the higher radio luminosities observed within this sample, as mergers have been found to generate the most luminous and radio-loud AGN \citep{Treister2012, Chiaberge2015}.

It thus appears as though the variations seen in AGN fraction are reflected by their radio luminosity in the same spatial regions. The AGN fraction is at its highest in the \textit{inner} region, and these AGN also appear to be the most radio luminous; whereas the AGN fraction is at its lowest in the \textit{intermediate} region, which also appears to hold the highest fraction of low luminosity AGN; finally, the \textit{outer} region, which has the \say{middling} galaxy fraction extremum, also appears to host the \say{middling} level of AGN activity across the luminosity range.

However, we do not see variations in the $i$-band stellar masses of the host galaxies of AGN in the \textit{intermediate} and \textit{outer} regions of galaxy clusters. The \textit{inner} region displays a clear increase in the number of massive host galaxies, which can be attributed to the presence of BCGs near the centre of most clusters \citep{Best2007}, but the mass distributions of AGN host galaxies in both the \textit{intermediate} and \textit{outer} regions are almost identical, suggesting that the fractional \say{wiggle} seen in \cref{fig:AGN_frac,fig:AGN_frac_AI} cannot be attributed to a difference in the host stellar mass.

The features seen and discussed here are all relatively subtle effects, which may be the reason why previous studies have found conflicting or inconclusive results. The combination of both AGN being quite rare astronomical objects, and the variation from the field fraction being only $\sim 20 - 25\%$, means that it would be easy to miss without a large population of clusters, as used here.

To conclude, we have determined that there is statistically significant variation in the fraction of radio AGN with respect to cluster-centric radius, the shape of which was predicted in a recent simulation paper by \citet{Rihtarsic2023}. The similarity of the host galaxy stellar masses in the \textit{intermediate} and \textit{outer} regions suggests that the differences in AGN fraction between these regions are not driven by their host galaxy properties. However, the variation in radio luminosity seen between all three regions suggests that AGN fraction isn't the only difference in these radial windows - AGN activity itself is distinct too. The combination of these results suggest that the effects seen are driven by different environmental mechanisms, depending on the AGN's location with respect to the nearest cluster, which cause them to oscillate between activity and inactivity independently of their host galaxy's stellar mass. In the case of AGN activity in the proximity of galaxy clusters - nurture wins out over nature.

\section*{Acknowledgements}

KdV, NAH and MRM acknowledge support from the UK Science and Technology Facilities Council (STFC) under grant ST/X000982/1.
All authors are grateful for the use of data from LOFAR, the LOw Frequency ARray, and SDSS, the Sloan Digital Sky Survey.
This research made use of Astropy, a community-developed core Python package for astronomy \citep{astropy:2013, astropy:2018} hosted at \url{http://www.astropy.org/}, of MATPLOTLIB \citep{Hunter:2007}, of Plotly \citep{plotly}, and of TOPCAT \citep{Taylor:2005}.

\section*{Data Availability}
All of the individual catalogues used in this paper are pubicly available. The LoTSS DR2 data can be found at \url{https://www.lofar-surveys.org/releases.html}, the SDSS DR16 data can be found at \url{https://www.sdss4.org/dr16/spectro/spectro_access/}, the MPA-JHU data can be found at \url{https://wwwmpa.mpa-garching.mpg.de/SDSS/DR7/}, and the allWISE data can be found at \url{https://irsa.ipac.caltech.edu/cgi-bin/Gator/nph-scan?mission=irsa&submit=Select&projshort=WISE}. The cluster catalogues by \citet{Wen2012} and \citet{Wen2015} are also publicly available and are associated with the referenced papers. For access to the cluster-matched data compiled from these catalogues, please contact KdV.



\bibliographystyle{mnras}
\bibliography{bibliography} 





\bsp	
\label{lastpage}
\end{document}